\documentclass[twocolumn,showpacs,amsmath,amssymb,pre,aps]{revtex4}   
\usepackage{graphicx}
\usepackage{dcolumn}
\usepackage{bm}
\usepackage{color}

\newif\ifgraph

\graphtrue             %

\begin{document}
\title{
Pattern formation in vortex matter with pinning and frustrated inter-vortex interactions
}
\draft

\author{
H.~J.~Zhao$^{1,2}$, V.~R.~Misko$^{2,3}$, J.~Tempere$^{2,4}$, and F.~Nori$^{3,5}$
}

\affiliation{$^1$
Department of Physics, Key Laboratory of MEMS of the Ministry of Education, Southeast University, Nanjing 211189, China
}
\affiliation{$^{2}$
TQC, Universiteit Antwerpen, Universiteitsplein 1,
B-2610 Antwerpen, Belgium
}
\affiliation{$^3$ 
CEMS, RIKEN, Saitama, 351-0198, Japan
}
\affiliation{$^{4}$
Lyman Laboratory of Physics, Harvard University, USA
}
\affiliation{
$^5$ Physics Department, University of Michigan, 
Ann Arbor, MI 48109-1040, USA
}

\date{\today}

\begin{abstract}
We investigate the effects related to vortex core deformations when vortices approach each other.
As a result of these vortex core deformations, the vortex-vortex interaction effectively acquires an attractive component leading to a variety of vortex patterns typical for systems with non-monotonic repulsive-attractive interaction, such as stripes, labyrinths, etc.
The core deformations are anisotropic and can induce frustration 
in the vortex-vortex interaction. 
In turn, this frustration has an impact on the resulting vortex patterns, which are analyzed in the presence of additional random pinning, as a function of the pinning strength. 
This analysis can be applicable to vortices in multiband superconductors or to vortices in Bose-Einstein condensates.
\end{abstract}
\maketitle

\section{Introduction}

Vortex-vortex interactions in superfluid atomic gases or type-II superconductors are purely repulsive, and can be described by a simple pairwise potential for two well-separated vortices. 
For dilute vortex matter, the description of the total interaction energy in terms of pairwise potentials is an approximation that is widely applicable and that describes many observed phenomena in both superfluids and superconductors, including the formation of the vortex lattice, vortex dynamics, vortex pinning, etc.~\cite{aftalion}. 
%
However, it must be kept in mind that this approximation leads to results that deviate from the full time-dependent Ginzburg-Landau (GL) results when 
(i) the vortex density becomes high (with cores in close vicinity of each other) or strongly inhomogeneous, or 
(ii) when the vortices move at a velocity comparable to the critical superfluid velocity and a transition from vortex channels to phase slip lines is possible. 
Despite these shortcomings, molecular dynamics simulations using pairwise potentials have been used very successfully to describe, 
for example, 
the vortex ratchet effect~\cite{Wambaugh,vvmratchet,linratchet}
or quasiperiodic vortex structures~\cite{wepenprl,wepenprb,pendieterprl,pensilhanekapl,penvvmprl}.  

Due to their spatial extension, the vortex-vortex interaction acquires
a threshold that can be effectively described as a sum of a pure repulsive vortex-vortex interaction potential and an additional attractive term.
This situation is 
similar 
to the earlier-studied case of multiscale vortex-vortex interactions when two or more purely repulsive potentials characterized by different length-scales result in the appearance of an attractive component in the resulting inter-vortex interaction (see, e.g., \cite{babaevPRB2014}).
The same idea stands behind the simple interpretation of the origin of an attractive interaction in two- and multi-band superconductors~\cite{Babaev2005,VVM2009,Nishio2010,Babaev2010,silhanek}, 
where different bands are characterized by different sets of the characteristics lengths, the coherence length, $\xi_{i}$, and the magnetic field penetration depth, $\lambda_{i}$. 
As a result, 
different bands have different GL parameters $\kappa_{i}$ that define the lengthscales of the vortex-vortex interaction. 
Furthermore, in the 
special case of a two-band superconductor 
(called ``type-1.5 superconductor''~\cite{VVM2009}), like MgB$_{2}$, two different GL parameters, 
corresponding to the different bands, 
can be either smaller or larger than the dual point, 
$\kappa_{\pi}<1/\sqrt{2}$ (type-I) and 
$\kappa_{\sigma}>1/\sqrt{2}$ (type-II), 
thus leading straightforward to a non-monotonic repulsive-attractive intervortex interaction~\cite{Babaev2005,VVM2009,Nishio2010}. 
Another example of a vortex system that acquires an attractive term in the intervortex interaction is a layered superconductor in tilted magnetic field~\cite{Berezin,BuzdinPRL,BuzdinPRB}. 
Vortices become anisotropic, due to the elongation in the direction of the field tilting, and interact attractively which may lead to the formation of stable vortex complexes~\cite{BuzdinPRB}. 

In this context, it is also worth mentioning so-called ``low-$\kappa$'' superconductors, i.e., materials with $\kappa \gtrsim 1/\sqrt{2}$ (called ``low-$\kappa$'' as opposite to the case of $\kappa \gg 1$~\cite{Auer}). 
The detailed calculations of the free energy of the vortex state 
in type-I and type-II superconductors 
first have been carried out decades ago
~\cite{eilenberger,Jacobs,brandt,klein}. 
It was also shown that 
materials 
with $\kappa$ in a very narrow range close to $\kappa \gtrsim 1/\sqrt{2}$
revealed attraction between vortices
(this narrow region close to the phase transition to type-I superconductivity was called ``type-II/1''~\cite{klein}, to distinguish it with type-II superconductivity). 
Recently, the interest to low-$\kappa$ superconductors has been renewed thanks to the advances in the studies of new materials and visualization techniques. 
Thus recent experiments with ZrB$_{12}$ and LuB$_{12}$ with $\kappa \gtrsim 1/\sqrt{2}$~\cite{sluchanko,ge} revealed the earlier theoretically predicted so-called intermediate mixed phase (IMP) [or intermediate mixed state (IMS)] which is a combination of the mixed phase and the Meissner phase. 
These experiments confirm that the intervortex interaction in the IMP is repulsive-attractive, and they allow one to analyze the transition to the type-I superconductivity where vortices are attractive (see, e.g., the recent experiment~\cite{prozorov}). 
Note that the appearance of the attraction between vortices in the IMP (i.e., in the vicinity to the dual point) is related to the onset of the {\it overlap} of the vortex cores. 
This provides a direct link of our model (described below) to low-$\kappa$ superconductors. 
In addition,
the appearance of an effective attractive term has been recently demonstrated also for
non-pairwise vortex-vortex interaction \cite{babaevPRB2015}.

In turn, systems interacting via repulsive-attractive potential (in particular, of Lennard-Jones type) were extensively studied in physics and were shown to result in a variety of non-trivial patterns, including stripes, labyrinths, lattices with voids, etc.
\cite{Vedmedenko2007,Ball1999}.
For vortices interacting via non-monotonic repulsive-attractive interactions, vortex pattern formation has been analyzed
\cite{FNsci97,FNprb97,FNprl98,FNprl99,kagome,Zhao2012NJP,Zhao2013PRE,babaevPRB2014,babaevPRB2015}, using various models.
In particular, vortex clusters, stripes, labyrinths, deformed lattices, and lattices with voids were found \cite{Zhao2012NJP}.
In addition, systems with non-monotonic interaction were shown to display unusual dynamics, such as size-selective dynamical cluster formation and re-orientation of longitudinal stripes to transverse stripes \cite{Zhao2013PRE}.
Some of the obtained static patterns were employed to explain the observed vortex patterns in superconductors, either in two-band materials, like MgB$_{2}$, 
or in low-$\kappa$ superconductors. 

Let us take a more careful look at the similarities and differences between the calculated regular vortex patterns (see, e.g.,
\cite{Zhao2012NJP,Zhao2013PRE,babaevPRB2014,babaevPRB2015,Xu}) and the observed vortex patterns in multi-band superconductors \cite{VVM2009,Nishio2010,silhanek}.
The most prominent features that the numerical simulations reproduce are the formation of vortex clusters and vortex stripes.
These are at the same time the most generic types of patterns derived from non-monotonic interactions \cite{Vedmedenko2007,Ball1999}.
Despite these basic similarities, the measured vortex patterns are much less perfect: e.g., less ordered broken stripes \cite{VVM2009,Nishio2010,silhanek} or 
chains of dimers rather than regular stripes \cite{bending}.
Clearly, these 
discrepancies 
require improvement of the employed theoretical approaches in order to reach a better understanding of the factors contributing to the vortex pattern formation.

In this work, we analyze effects related to
(i) a short-range effective attraction in the overall repulsive intervortex interaction, in presence of random pinning, and 
(ii) frustration in the vortex-vortex interaction, combined with random pinning.
The latter effect, 
random pinning, 
is rather obvious: pinning is inevitably present in superconductors (although it is less evident in case of vortices in 
Bose-Einstein condensates (BEC)~\cite{bookbec}), 
and it clearly has an impact on vortex pattern formation. 
The 
appearance of frustration in the intervortex interaction 
can be understood from the fact that, when deformed, vortex cores {\it elongate} in the direction of the closest neighbor resulting in an anisotropy and thus breaking the symmetry of the interaction in the system~\cite{jtprl2012,jtpc2012}. 
As a result, the interaction of the vortex with elongated core with a {\it second} closest neighbor will depend on the {\it orientation} of the vortex core with respect to the direction to that second closest neighbor. 
Clearly that in the ideal case of two equally-close neighbors (like, e.g., in case of antiferromagnetically-interacting spins placed on vertices of 2D polygons with odd number of vertices or on vertices of 3D tetrahedra) the vortex-vortex interaction appears to be {\it frustrated}: the chosen vortex should ``decide'' whether to elongate in the direction of the first neighbor or in the direction of the second neighbor.
Geometrical frustration has been extensively studied in physics \cite{frankel,sadoc} including condensed matter physics \cite{nelson},
liquids and glasses \cite{nelsonprb83,steinhardt}, and superconducting vortices in various artificial pinning arrays (APS) \cite{kagome,wepenprl,wepenprb,weht}.
Here we will analyze the effect of frustration that appears in the vortex-vortex interaction due to deformations of vortex cores being perturbed by close neighbors.

The paper is organized as follows. 
The model is introduced in Sec.~II. 
In Sec.~III, we analyze the effects related to core deformations 
without pinning and in the presence of pinning, assuming isotropic 
case not leading to frustration. 
The calculated vortex patterns are compared to the experimental 
images. 
Effects related to anisotropy and frustration in the vortex-vortex interaction are discussed in Sec.~IV, and a comparison of the 
calculated vortex patterns to the experimental patterns is presented. 
The conclusions of this work are summarized in Sec.~V. 

\section{Model} 

We model a 3D column, infinitely long in the $z$-direction,
by a 2D (in the $xy$-plane) square simulation cell
with periodic boundary conditions.
To study the dynamics of vortex motion,
we numerically integrate the overdamped equations of motion
(see, e.g., Refs.~\cite{wepenprl,wepenprb}):
\begin{equation}
\eta {\rm \bf v}_{i} \ = \ {\rm \bf f}_{i} \ = \ {\rm \bf f}_{i}^{vv} + {\rm \bf f}_{i}^{vp} + {\rm \bf f}_{i}^{T} + {\rm \bf f}_{i}^{d}.
\label{leq}
\end{equation}
Here
${\rm \bf f}_{i}$
is the total force per unit length acting on vortex
$i$,
${\rm \bf f}_{i}^{vv}$
and
${\rm \bf f}_{i}^{vp}$
are the forces due to vortex-vortex and vortex-pin interactions, respectively,
${\rm \bf f}_{i}^{T}$
is the thermal stochastic force,
and
${\rm \bf f}_{i}^{d}$
is the driving force;
${\rm \bf v}_{i}$
is the velocity, and
$\eta$ is the viscosity.
All the forces are expressed in units of
$
f_{0} = \Phi_{0}^{2} / 8 \pi^{2} \lambda^{3},
$
where $\Phi_{0} = hc/2e$, and
lengths (fields)
in units of
$\lambda$ ($\Phi_{0}/\lambda^{2}$).

\begin{figure}[btp]
\begin{center}
\includegraphics*[width=8.5cm]{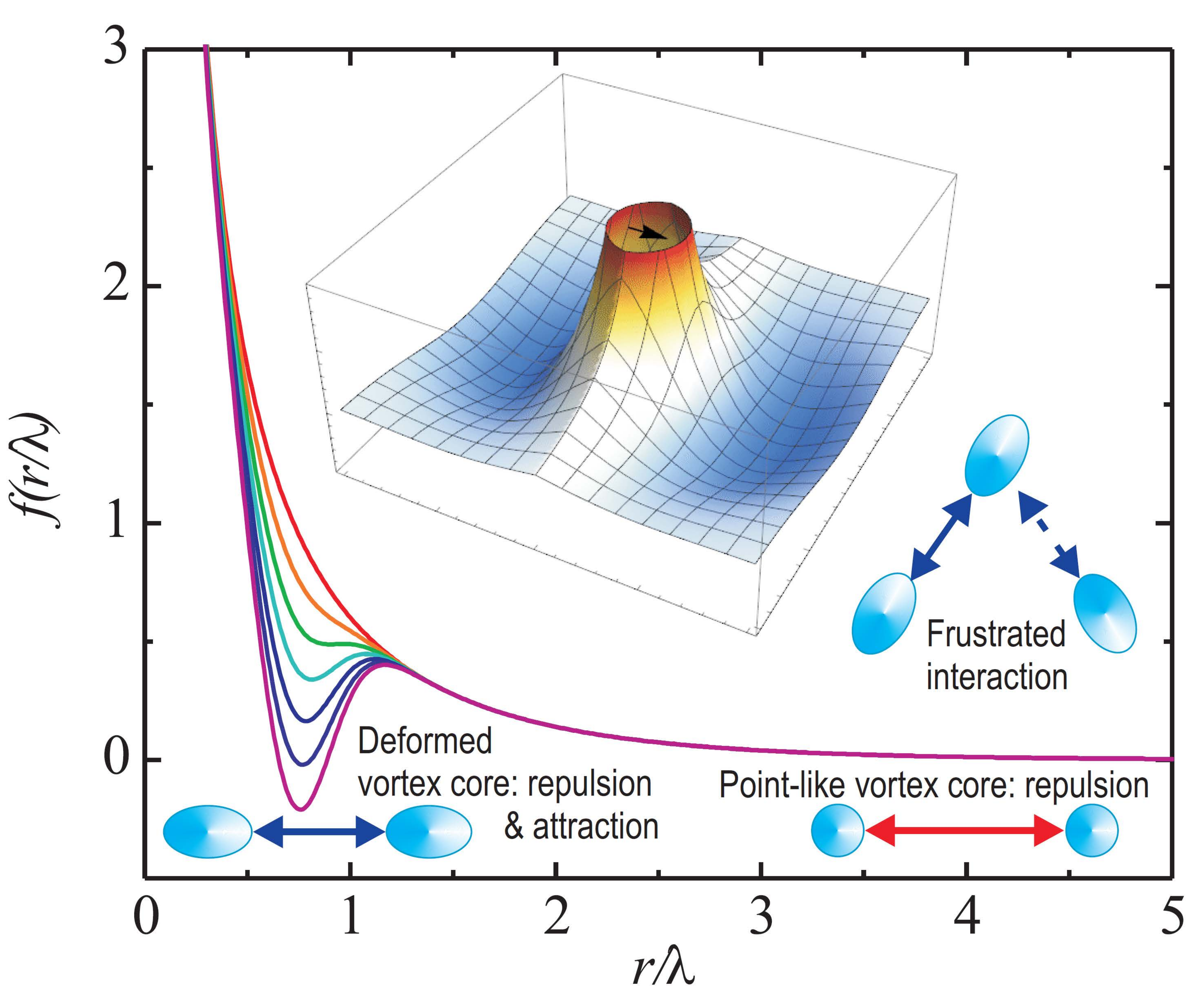}
\end{center}
\vspace{-0.5cm}
\caption{
Schematic plot of the model vortex-vortex interaction.
At large distances, vortices characterized by unperturbed circular-symmetric cores (as shown in the righ-hand lower inset), are repelled by the repulsive inter-vortex interaction force (red arrow in the inset).
Eventual vortex core deformations at short distances (as shown in the left-hand lower inset) give rise to the appearance of effectively multiscale vortex-vortex interaction which, in turn, leads to an additional attractive term in the vortex-vortex interaction.
The degree of the core deformation and the corresponding strength of the vortex-vortex attraction force is controlled by the parameter $\beta$. 
In the plot, the curves are shown for 
$\beta=0$, 0.2, 0.4, 0.6, 0.8, 1.0, and 1.2, 
$\gamma=14$, and $r_{0}=0.7$. 
The left-hand upper inset shows the profile of the vortex-vortex 
interaction force in the $xy$-plane, for an anisotropic 
interaction, 
where the direction of the vector $\hat{\rm \bf r}_{ik}$ 
[see the notations of Eq.~(\ref{fvv2})] is shown by the black arrow. 
The right-hand upper inset illustrates the effect of frustration in the inter-vortex interaction when two close-neighbor vortices with elongated cores are equally separated from a given vortex.
}
\label{int}
\end{figure}

The force due to the interaction of the $i$-th vortex with other vortices (see, e.g., Refs.~\onlinecite{FNprl94,FNsci97,FNprb97,wepenprl,wepenprb}) is: 
\begin{equation}
{\rm \bf f}_{i}^{vv} \ = \ \sum\limits_{j}^{N_{v}} \ f_{0} \ K_{1} \!
\left( \frac{ \mid {\rm \bf r}_{i} - {\rm \bf r}_{j} \mid }{\lambda} \right)
\hat{\rm \bf r}_{ij} \; ,
\label{fvv}
\end{equation}
where
$N_{v}$
is the number of vortices,
$K_{1}$
is a first-order modified Bessel function,
and
$\hat{\rm \bf r}_{ij} = ( {\rm \bf r}_{i} - {\rm \bf r}_{j} )
/ \mid {\rm \bf r}_{i} - {\rm \bf r}_{j} \mid.$
To study the effects related to vortex-core deformations and frustration, we modify Eq.~(\ref{fvv}) by introducing an additional attractive term in the form of a Gaussian 
(cp. Ref.~\onlinecite{babaevPRB2014} for multiscale inter-vortex interaction),
\begin{eqnarray} 
{\rm \bf f}_{i}^{vv} & = & \sum\limits_{j}^{N_{v}} \ f_{0} \Biggl\{ \ K_{1} \! \left( \frac{ \mid {\rm \bf r}_{i} - {\rm \bf r}_{j} \mid }{\lambda} \right) \nonumber \\
& - & \beta 
\Omega
\exp\left[{- \gamma\left(\frac{ \mid {\rm \bf r}_{i} - {\rm \bf r}_{j} \mid }{\lambda}- r_{0}\right)^{2}}\right]
\Biggr\}
\hat{\rm \bf r}_{ij} \; ,
\label{fvv2}
\end{eqnarray} 
where the model parameters $\gamma$ and $r_{0}$ are fixed 
($\gamma=14$, $r_{0}=0.7$),
and we vary the attraction strength $\beta$ in our simulations. 
The coefficient $\Omega$ is $\Omega=1$ for an isotropic interaction, 
and 
$\Omega = \ \mid (\hat{\rm \bf r}_{ij} \cdot \hat{\rm \bf r}_{ik}) \mid$, 
where 
$(... \cdot ...)$ denotes a scalar product of two vectors, and 
$\hat{\rm \bf r}_{ik} = ( {\rm \bf r}_{i} - {\rm \bf r}_{k} )
/ \mid {\rm \bf r}_{i} - {\rm \bf r}_{k} \mid$ 
is a unit vector along the direction connecting vortex $i$ and 
the 
{\it closest-neighbor} vortex $k$, 
in the anisotropic case when frustration in the vortex-vortex interaction is taken into account. 
The modified vortex-vortex model potential is illustrated in Fig.~\ref{int}.

Vortex pinning is modeled by short-range parabolic potential wells
located at positions
${\rm \bf r}_{k}^{(p)}$.
The pinning force is
\begin{equation}
{\rm \bf f}_{i}^{vp} = \sum\limits_{k}^{N_{p}} \left( \frac{f_{p}}{r_{p}} \right)
\mid {\rm \bf r}_{i} - {\rm \bf r}_{k}^{(p)} \mid
\Theta \!
\left(
\frac{r_{p} - \mid {\rm \bf r}_{i} - {\rm \bf r}_{k}^{(p)} \mid}{\lambda}
\right)
\hat{\rm \bf r}_{ik}^{(p)},
\label{fvp}
\end{equation}
where
$N_{p}$
is the number of pinning sites,
$f_{p}$
is the maximum pinning force of each potential well,
$r_{p}$
is the range of the pinning potential,
$\Theta$
is the Heaviside step function,
and
$\hat{\rm \bf r}_{ik}^{(p)} = ( {\rm \bf r}_{i} - {\rm \bf r}_{k}^{(p)} )
/ \mid {\rm \bf r}_{i} - {\rm \bf r}_{k}^{(p)} \mid.$

The temperature contribution ${\rm \bf f}_{i}^{T}$
is represented by a stochastic
term obeying the following conditions:
\begin{equation}
\langle f_{i}^{T}(t) \rangle = 0
\end{equation}
and
\begin{equation}
\langle f_{i}^{T}(t)f_{j}^{T}(t^{\prime}) \rangle = 2 \,  \eta \,  k_{B} \,  T \,  \delta_{ij} \,  \delta(t-t^{\prime}).
\end{equation}

To obtain the ground state of a system of vortices,
the system starts at
some non-zero value of the ``temperature'' and gradually decrease
it to zero, i.e., we perform a simulated-annealing simulation.
This procedure mimics the annealing procedure in field-cooled experiments.

\section{Attractive component and pinning}

First we consider the effect of a short-range attractive component in the vortex-vortex interaction. 
This may arise from the relaxation deformation of vortex cores when two vortices come in close proximity~\cite{coredeformation}, or from the multiband nature of the underlying superconductor or superfluid. 

Above, in Sec.~I, we discussed the origin of the attractive intervortex interaction in the case of multi-band superconductors. 
The effective attraction, as explained above, arises from different lengthscales for the different bands~\cite{babaevPRB2014}, or from the different signs of the interaction in the different bands, in the case of the type-1.5 superconductors~\cite{Babaev2005,VVM2009,Nishio2010}. 
Note that the term ``effective'' here means that the interaction force between two vortices might not necessarily change the sign or even have a local minimum, but only become {\it lower} in absolute value for some $r$ (see Fig.~\ref{int}), i.e., the force becomes ``less repulsive'' as compared to the bare expression, Eq.~(\ref{fvv}). 

In Sec.~I, we also discussed the relation between our model and low-$\kappa$ superconductors. 
In the vicinity to the dual point, where $\kappa \gtrsim 1/\sqrt{2}$, the size of the vortex core is nearly the same as its magnetic core. 
This provides a gentle balance between the repulsive and attractive contributions and results in a weak repulsion near the dual point (but still in the type-II/1 region) and, at the same time, makes this state unstable with respect to the phase transition to the type-I state. 
It is clear that fluctuations of any nature, e.g., due to the motion of vortices or due to a sudden trapping of two neighboring vortices by a pinning site, may {\it locally} induce the transition to the type-I state. 
This occurs due to partial {\it overlapping} of the vortex cores (the source of vortex attraction in type-I superconductors)~\cite{overlapping}. 
The vortex-core overlapping may also result from sudden deformations or anisotropy of vortex cores when two vortices are in close vicinity. 
This results in the appearance of an attractive component in the intervortex interaction in low-$\kappa$ superconductors even for single-band materials characterized by a repulsive intervortex interaction far from the type-II/1 to type-I phase transition. 
Above, in Sec.~I, we mentioned that vortex anisotropy and intervortex attraction may also result from magnetic field tilting in layered superconductors~\cite{Berezin,BuzdinPRL,BuzdinPRB}. 
However, this mechanism leads to the global anisotropy (i.e., all the vortices become elongated and acquire the attractive interaction) in the direction of the field tilting. 
This situation can hardly result in frustration of the intervortex interaction and thus is not considred in the present work. 
Instead, we focus on systems where core deformations may occur locally and thus lead to frustration in the intervortex interaction. 
These systems include, but are not limited to, multi-band superconductors or low-$\kappa$ superconductors where core deformations at certain conditions, as described above, may lead to or enhance the intervortex attraction. 

One more interesting recent example of a physical system, that can be treated within our model, is a superconducting device that allows for the observation of the transition from type-II and type-I behavior, in one sample~\cite{gladilin}. 
The sample has a shape of a superconducting wedge with a varying thickness that provides a smooth transition from the effective type-II superconductor (i.e., with $\kappa_{eff}>1/\sqrt{2}$) to the material with effective type-I parameters ($\kappa_{eff}<1/\sqrt{2}$). 
Using the time-dependent GL equations, the authors showed that current-driven flux patterns in this device undergo the transition from the Abrikosov vortex lattice to the mixed state in the type-I supercondutor via a series of transient vortex-molecule or/and giant-vortex states. 

\subsection{Zero pinning case}

\begin{figure}[btp]
\begin{center}
\includegraphics*[width=8.5cm]{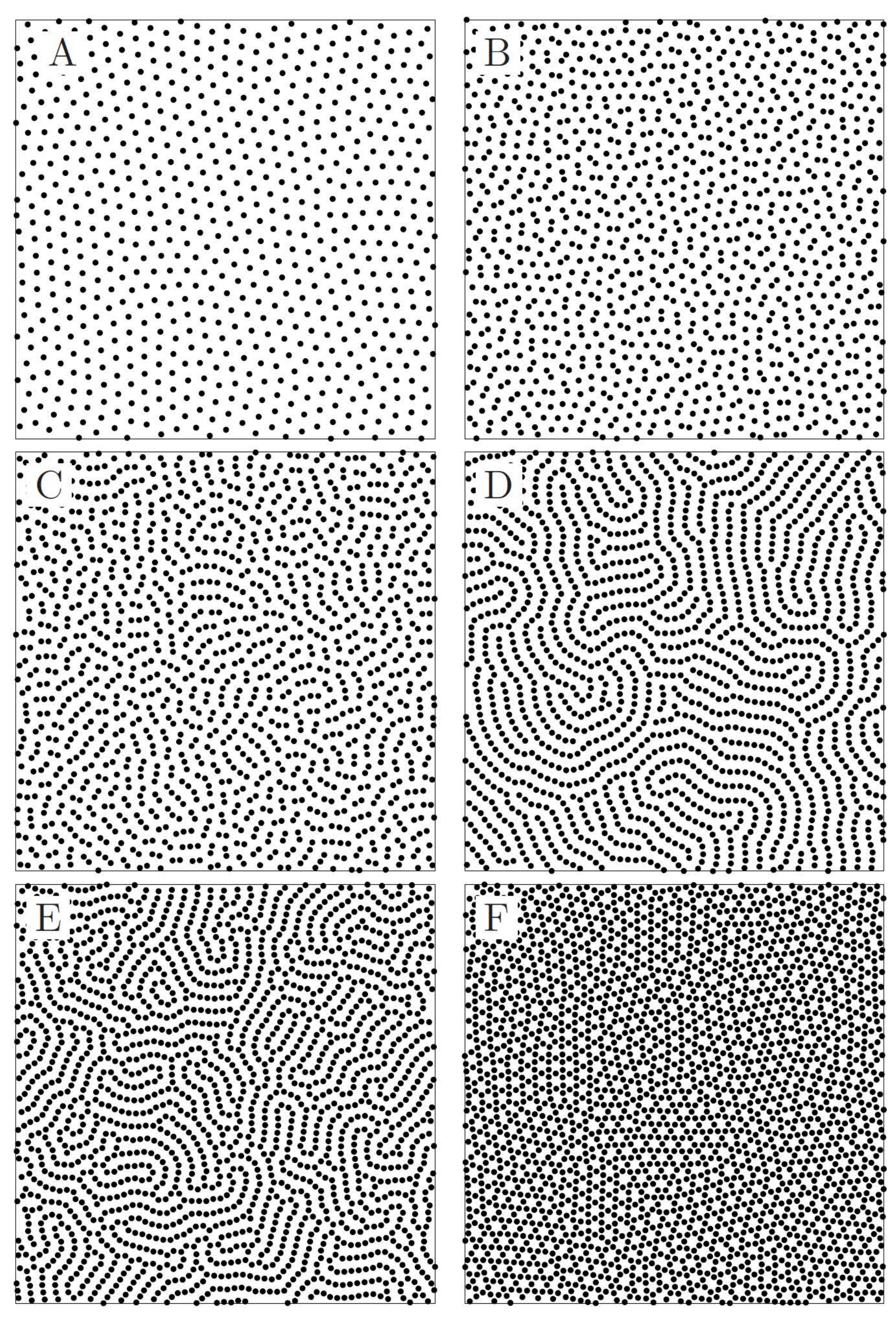}
\end{center}
\vspace{-0.5cm}
\caption{
Vortex patterns for $\beta=0.6$
in the absence of pinning
and
for varying number of vortices
per simulation cell:
$N_{v}=1600$ (A),
$N_{v}=2500$ (B),
$N_{v}=3025$ (C),
$N_{v}=3600$ (D),
$N_{v}=4225$ (E), and
$N_{v}=5625$ (F).
}
\label{nfnpb06n}
\end{figure}

In our model,
the attractive component is characterized by a non-zero parameter $\beta$ [Eq.~(\ref{fvv2})]. 
Here the pinning strength is set to zero, and we first focus on the effects related to non-zero $\beta$ and changing vortex density.
The results of simulations for $\beta=0.6$ and varying vortex density, or number of vortices per simulation region $N_{v}$, are presented in Fig.~\ref{nfnpb06n}.

For low vortex densities, when the average intervortex distance is
larger than the characteristic distance at which the attractive component in the vortex-vortex interaction comes into play, vortices arrange themselves in a hexagonal (Abrikosov) lattice [see Fig.~\ref{nfnpb06n}(A) for $N_{v}=1600$].
Increasing the vortex density above this limit (e.g., $N_{v}=2500$) is characterized by the appearance of dimers [Fig.~\ref{nfnpb06n}(B)], due to the symmetry breaking induced by the attractive interaction.
The hexagonal lattice is completely destroyed.
Instead, there is a disordered mixture (liquid phase) of single vortices and dimers showing an onset of stripe formation.
Next, for $N_{v}=3025$, the dimers develop to short stripes, as shown in [Fig.~\ref{nfnpb06n}(C)] which further evolve to long branching stripes [see Fig.~\ref{nfnpb06n}(D) for $N_{v}=3600$].
The stripes become denser with further increasing the vortex density forming labyrinths with some closed chains as shown in Fig.~\ref{nfnpb06n}(E) for $N_{v}=4225$.
Even higher vortex densities, e.g., $N_{v}=5625$ [Fig.~\ref{nfnpb06n}(F)], do not support one-dimensional (1D) stripes anymore, and the system undergoes a transition to kagom\'{e} lattice.

\begin{figure}[btp]
\begin{center}
\includegraphics*[width=8.5cm]{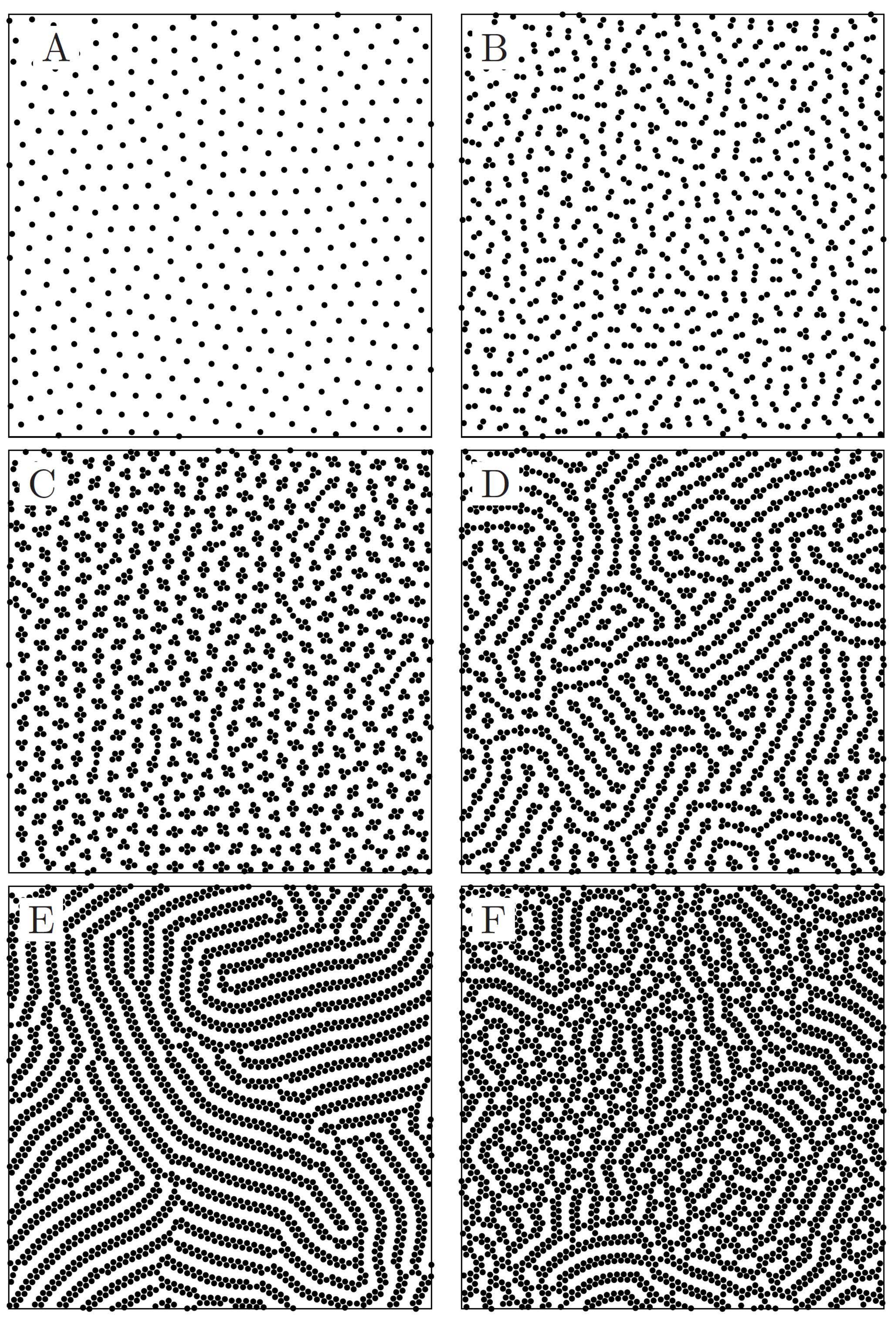}
\end{center}
\vspace{-0.5cm}
\caption{
Vortex patterns for $\beta=1.2$
in the absence of pinning
and
for varying number of vortices
per simulation cell:
$N_{v}=900$ (A),
$N_{v}=2025$ (B),
$N_{v}=3600$ (C),
$N_{v}=4225$ (D),
$N_{v}=5625$ (E), and
$N_{v}=6400$ (F).
}
\label{nfnpb12n}
\end{figure}

A set of vortex patterns for a larger value of $\beta=1.2$ and varying vortex density are presented in Fig.~\ref{nfnpb12n}.

As for the above case of smaller value of $\beta$, vortices form hexagonal lattice for low vortex densities [see Fig.~\ref{nfnpb12n}(A) for $N_{v}=900$].
When increasing the vortex density (e.g., $N_{v}=2025$) the symmetry of the system becomes broken, and the vortex pattern is represented by a disordered mixture (liquid) of dimers, 1D straight trimers and 2D trimers as shown in Fig.~\ref{nfnpb12n}(B).
Thus, unlike in case of a weaker attraction ($\beta=0.6$), the strong attraction characterized by $\beta=1.2$ facilitates the formation 
of not only 
1D but also 2D vortex clusters or chains.
This can be further seen in Fig.~\ref{nfnpb12n}(C) and Fig.~\ref{nfnpb12n}(D) for $N_{v}=3600$ and $N_{v}=4225$, correspondingly, when the vortices form first small 2D clusters consisting of three or four vortices which first interconnect by 1D chains (C) and then form 1D-2D chains (D).
At high enough vortex density, $N_{v}=5625$, the chains becomes long and purely 2D [Fig.~\ref{nfnpb12n}(E)].
For even higher vortex density, $N_{v}=6400$, the 2D chains break apart and interconnect forming a mixed state of interconnected 2D chains and 2D kagom\'{e} lattices [Fig.~\ref{nfnpb12n}(F)].

\subsection{Effect of pinning}

As one can see from the above results, when the attractive term related to the core deformation is taken into account, a vortex system undergoes a series of phase transitions driven by the attractive component in the vortex-vortex interaction and increasing vortex density.
In particular, the following phases (or ``morphologies'') and their sequence were revealed with increasing vortex density:
(i) a hexagonal vortex lattice,
(ii) a liquid of vortex dimers,
(iii) short stripes,
(iv) long stripes,
(v) interconnected stripes and labyrinths, and
(vi) kagom\'{e} lattices.
All the revealed patterns are rather ``perfect'' for commensurate vortex densities.
However, with the exception of the hexagonal vortex lattice, which is the most commonly observed and robust vortex structure in superconductors and Bose-Einstein condensates, none of the perfect 
patterns
typical for non-monotonic repulsive attractive interactions have been observed experimentally.
The main reason is, as we demonstrate below, that all these vortex patterns (except of the hexagonal lattice) are rather sensitive to imperfections in the system.
We analyze the role of imperfections on vortex patterns by introducing a random pinning and by varying the pinning strength $f_{p}$.

\begin{figure}[btp]
\begin{center}
\includegraphics*[width=8.5cm]{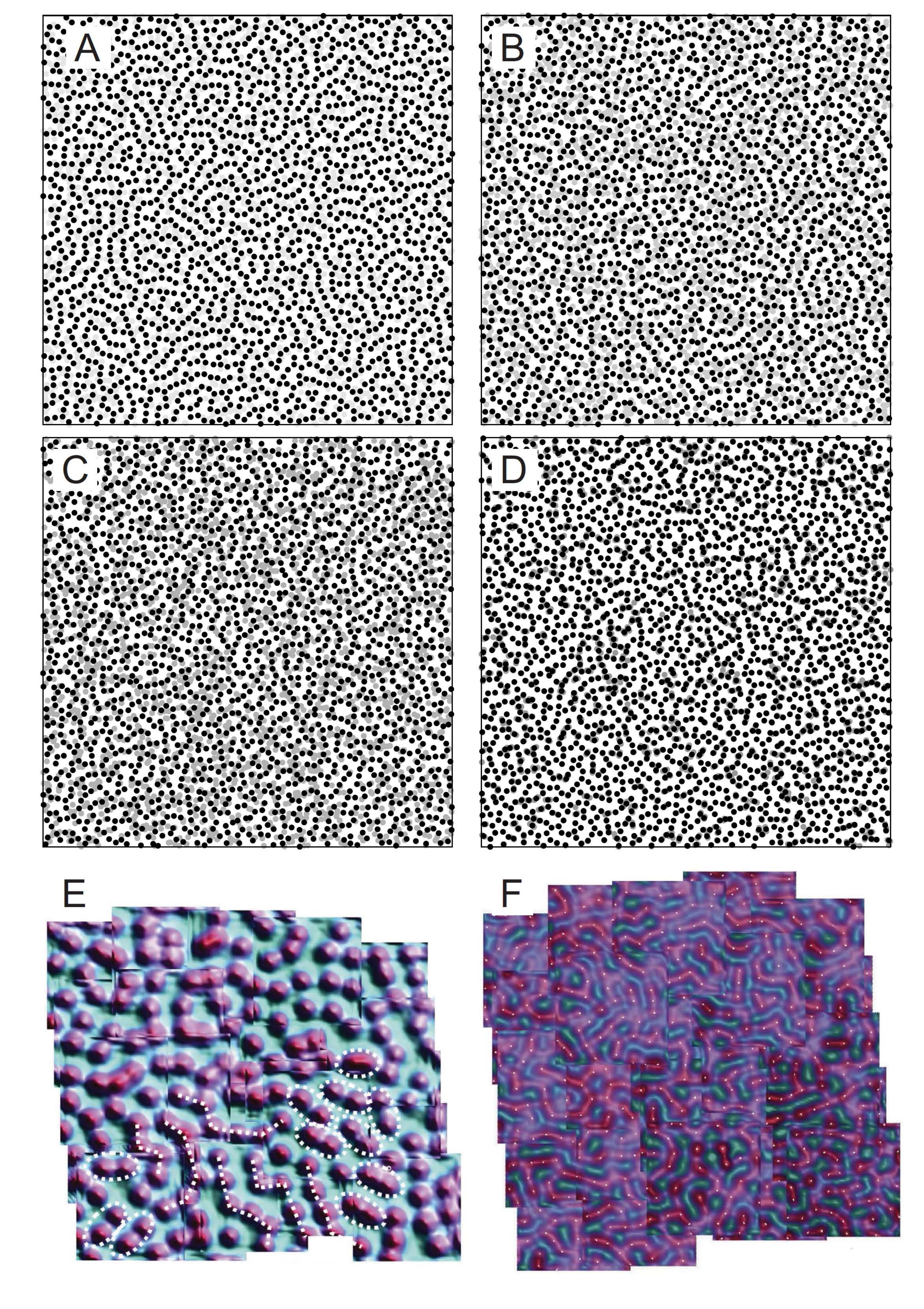}
\end{center}
\vspace{-0.5cm}
\caption{
Vortex patterns for $\beta=0.6$
and
for varying
pinning strength and
number of vortices
per simulation cell:
$f_{p}=0.3$, $N_{v}=4225$ (A),
$f_{p}=0.6$, $N_{v}=4225$ (B),
$f_{p}=0.9$, $N_{v}=4225$ (C), and
$f_{p}=1.2$, $N_{v}=4900$ (D).
Panels (E) and (F) show scanning Hall probe images ($\approx 50 \times 50 \ \mu$m$^{2}$) of the vortex distribution in a 160 nm thick superconducting MgB$_{2}$ film at T = 1.7 K and magnetic fields
of 1.7 Oe (E), 5 Oe (F) \cite{bending}.
}
\label{nfb06p}
\end{figure}

The effect of pinning is demonstrated in Fig.~\ref{nfb06p} for $\beta=0.6$.
As shown in Fig.~\ref{nfb06p}(A), even a weak pinning, $f_{p}=0.3$, strongly influences vortex patterns other than hexagonal lattice (a hexagonal lattice is rather robust to random pinning, as follows from our simulations (not shown) and from previous studies \cite{walter2009,walter2010}.
Thus, the initially well-ordered long stripes [see Fig.~\ref{nfnpb06n}(E) for $N_{v}=4225$] turn to rather irregular labyrinths when a weak pinning, $f_{p}=0.3$, is added to the system [Fig.~\ref{nfb06p}(A)].
Note that the morphology of the pattern is not changed.
These are still interconnected stripes (labyrinths) but they become less ordered and shorter in presence of pinning.
This indicates that the vortex-pin interaction is rather weak as compared to the vortex-vortex interaction, and the vortex stripes are pinned {\it collectively} \cite{walter2009,walter2010}.

What happens next, with increasing the pinning strength, the junctions between stripes break, as shown in Fig.~\ref{nfb06p}(B) for the same vortex density, $N_{v}=4225$ and $f_{p}=0.6$.
This can be understood if we compare Fig.~\ref{nfnpb06n}(C) and Fig.~\ref{nfnpb06n}(D) above.
The formation of junctions between stripes requires stronger intervortex attraction [which is achieved in Fig.~\ref{nfnpb06n}(D) by decreasing, as compared to (C), the average intervortex distance with increasing vortex density]. 
Correspondingly, these inter-stripe junctions appear to be easier to destroy by disorder.
For even stronger pinning, $f_{p}=0.9$ [Fig.~\ref{nfb06p}(C)], we observe a change in the morphology of the vortex pattern: not only vortex stripes become shorter and more disordered but we also observe a mixture of collective vortex pinning events (pinning of stripes) and {\it individual} vortex pinning \cite{walter2009,walter2010}.
Finally, for a very strong pinning force, $f_{p}=1.2$ [Fig.~\ref{nfb06p}(D)], practically all the vortices appear to be pinned by the pinning sites either individually or collectively, in the form of short stripes that fit into the pinning landscape. 

A comparison to typical experimental vortex images
[shown in panels (E) and (F)]
obtained using scanning Hall probe microscopy in a superconducting MgB$_{2}$ film at T = 1.7 K and magnetic fields of 1.7 Oe (E), 5 Oe (F) \cite{bending},
clearly indicates that the experimental patterns represent a mixture of individual disordered vortices and short stripes (E) or some longer stripes and individual vortices (F). 
This comparison allows us to identify the experimental images as the result of the interplay of two factors, the non-monotonic repulsive-attractive vortex-vortex interaction (which is due to vortex core deformations or the multi-band nature of MgB$_{2}$) {\it and} pinning in MgB$_{2}$ films.

\begin{figure}[btp]
\begin{center}
\includegraphics*[width=8.5cm]{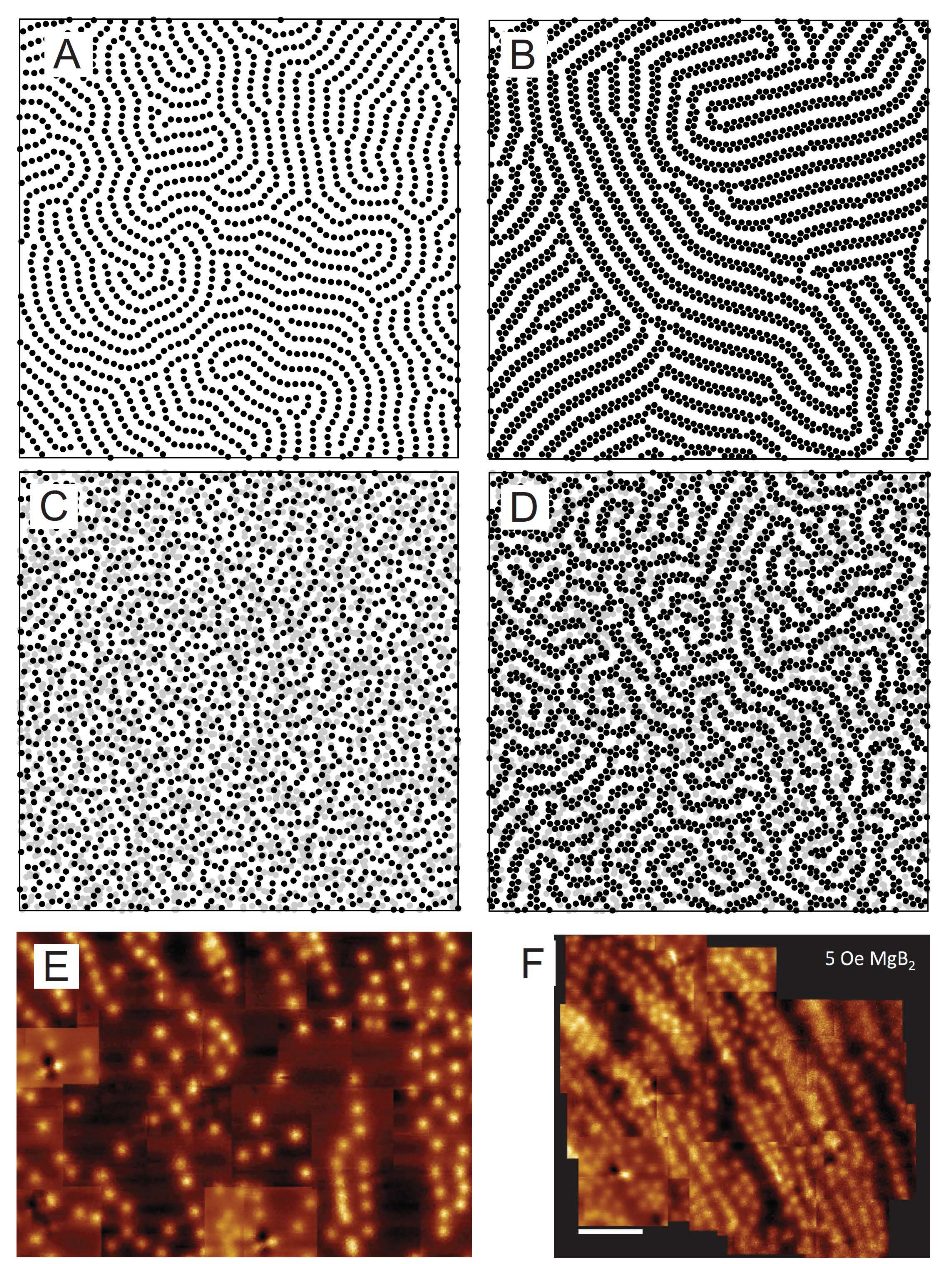}
\end{center}
\vspace{-0.5cm}
\caption{
Vortex patterns for $\beta=0.6$ and $\beta=1.2$,
and
for varying
pinning strength and
number of vortices
per simulation cell:
$\beta=0.6$, $f_{p}=0$, $N_{v}=3600$ (A),
$\beta=1.2$, $f_{p}=0$, $N_{v}=5625$ (B),
$\beta=0.6$, $f_{p}=0.6$, $N_{v}=3600$ (C), and
$\beta=1.2$, $f_{p}=0.6$, $N_{v}=5625$ (D).
Panels (E) and (F) show field-cooled images of vortex patterns in MgB$_{2}$ at 1 Oe (E) and 5 Oe (F) \cite{silhanek}.
}
\label{nfb12p}
\end{figure}

The effect of pinning is further analyzed in Fig.~\ref{nfb12p} for $\beta=0.6$ and $\beta=1.2$.
As unperturbed reference vortex patterns we consider long 1D stripes and long 2D stripes shown in Fig.~\ref{nfb12p}(A) and Fig.~\ref{nfb12p}(B), respectively. 
The same amount of disorder added to the system, $f_{p}=0.6$,
is shown to either break up the junctions of the long 1D stripes and shorten them [Fig.~\ref{nfb12p}(C)] or shorten and partially disorder long 2D stripes (which, however, remain predominantly 2D stripes, with inclusion of 1D elements, Fig.~\ref{nfb12p}(D).

Panels (E) and (F) of Fig.~\ref{nfb12p} show field-cooled images of vortex patterns in MgB$_{2}$ at 1 Oe (E) and 5 Oe (F) \cite{silhanek}.
The morphologies of the experimental images can be referred to as
either 1D stripes, branching 1D to 2D stripes, and disordered individual vortices (E), or 2D stripes, branching 1D to 2D stripes and disordered individual vortices (F).
This analysis of morphologies allows us to identify the experimental images as the result of the interplay of the repulsive-attractive vortex-vortex interaction and pinning in MgB$_{2}$ films.
(Note that the experimental images show somewhat elongated and curved 1D or 2D vortex stripes, which is a result of a particular pinning landscape in the measured samples \cite{silhanek}).

\begin{figure}[btp]
\begin{center}
\includegraphics*[width=8.5cm]{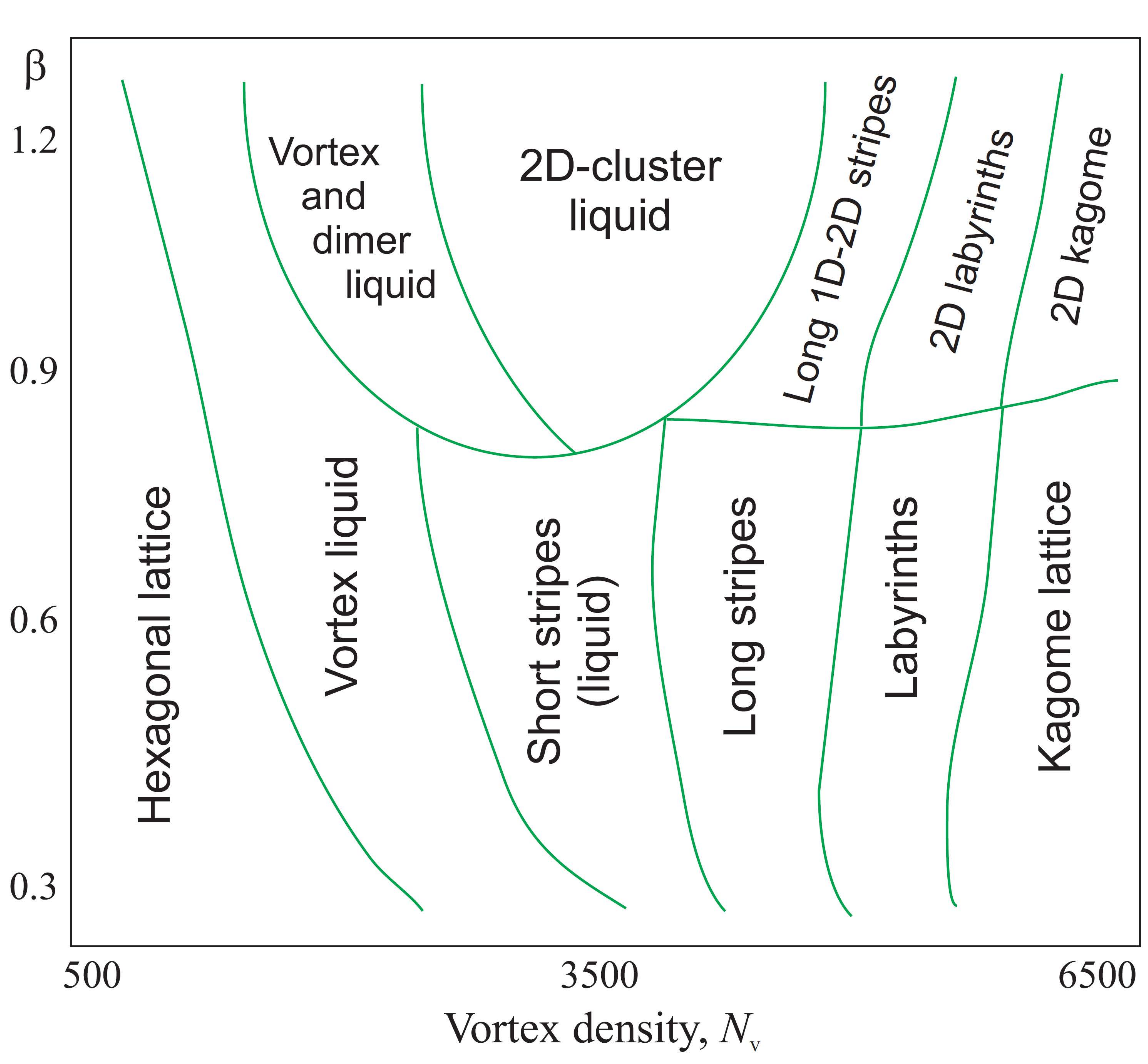}
\includegraphics*[width=8.5cm]{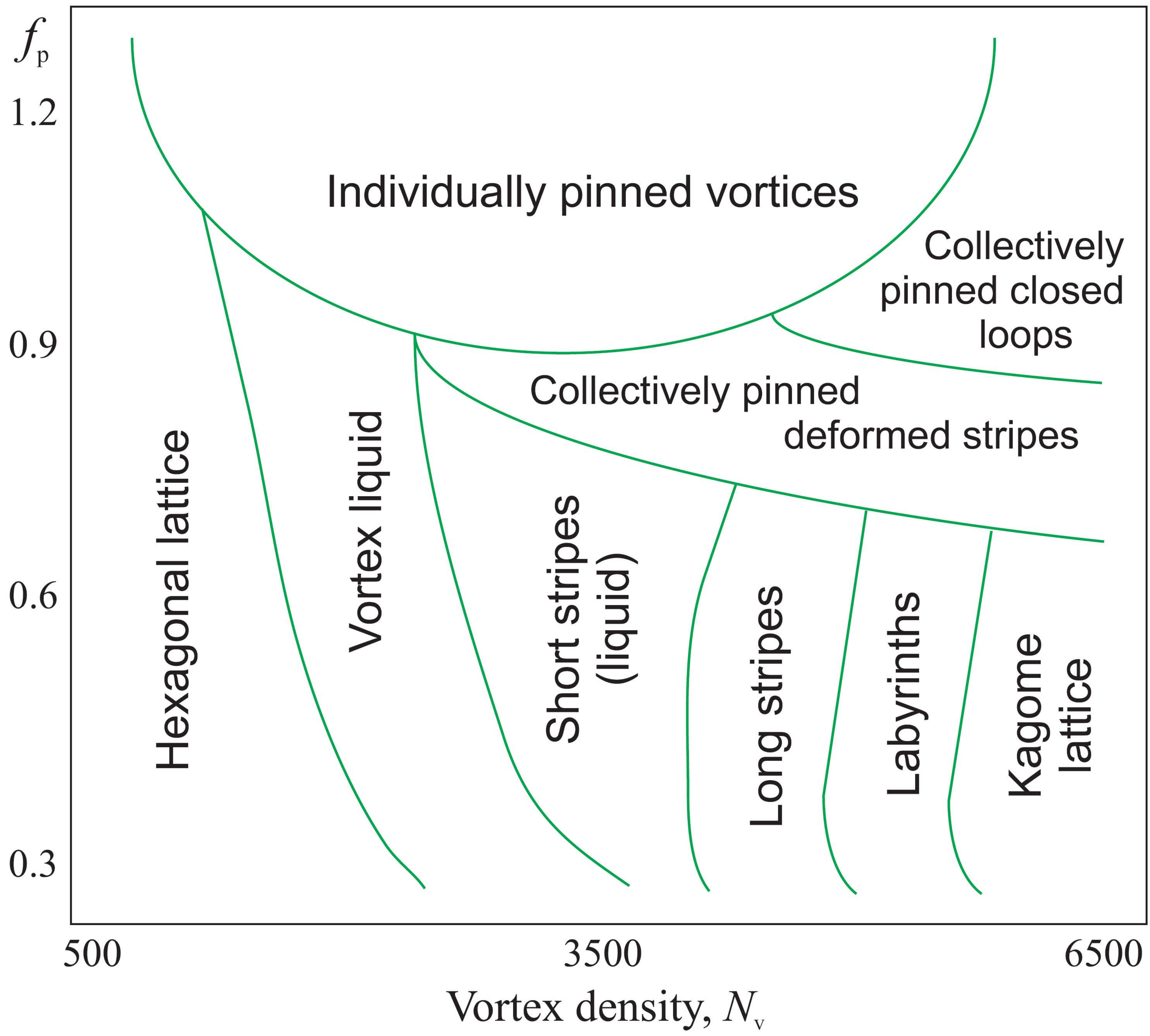}
\end{center}
\vspace{-0.5cm}
\caption{
These diagrams show the various morphologies of the vortex patterns that are encountered when varying the inter-vortex attraction force strength and the vortex density $N_{v}$ (upper panel), and when varying the random pinning strength $f_{p}$ versus the vortex density $N_{v}$ (lower panel). 
}
\label{phases}
\end{figure}

The calculated phases (or morphologies) of vortex patterns as a function of the vortex density (i.e., the number of vortices per simulation cell, $N_{v}$) revealed for varying parameters, the attraction strength $\beta$ and the pinning strength $f_{p}$, are shown in Fig.~\ref{phases}.

\section{Frustration in the vortex-vortex interaction}

Core deformations may lead to a non-monotonic interaction, but this non-monotonicity may have other sources such as the multiband nature of the underlying superconductor~\cite{VVM2009,Nishio2010,Babaev2005,Babaev2010}. 
However, core deformations can also alter the potential in a more fundamental way: Since the deformation can be anisotropic, they can introduce orientational frustration in the system. 
Here we modify our model by introducing an {\it orientational} order in the vortex-vortex interaction.
We assume that a vortex core elongates only in the direction of the {\it closest}-neighbor vortex, and therefore only this closest-neighbor vortex experiences the effective attraction to the chosen vortex as well as any vortex situated on the opposite side along the line connecting the interacting vortex pair.
Other vortices experience pure repulsive interaction from the chosen vortex.
In this situation (see the inset in Fig.~\ref{int}), it is possible that two neighbor vortices approach the chosen vortex at the same short distance when the chosen vortex should ``decide'', in what direction to elongate: either to the first or to the second neighbor vortex? 
This can lead to geometric frustration in the vortex-vortex interaction (although, in practice, there is always some small difference between the two short distances in numerical simulations that would eliminate frustration at the annealing stage). 
However, a more important expected consequence of the orientational deformation of the vortex core in the direction to the closest neighbor is a {\it trapping} of this neighbor vortex by the attractive potential and the formation of vortex dimers (and vortex stripes at higher densities).
As we demonstrate in our simulations, frustration in the vortex-vortex interaction manifests itself in the appearance of instability of vortex stripes with respect to their fragmentation into vortex dimers.
Indeed, fluctuations in the intervortex distance in a vortex stripe due to elastic deformations of the stripe will result in breaking the stripe apart in favor of vortex dimers.

\begin{figure}[btp]
\begin{center}
\includegraphics*[width=8.5cm]{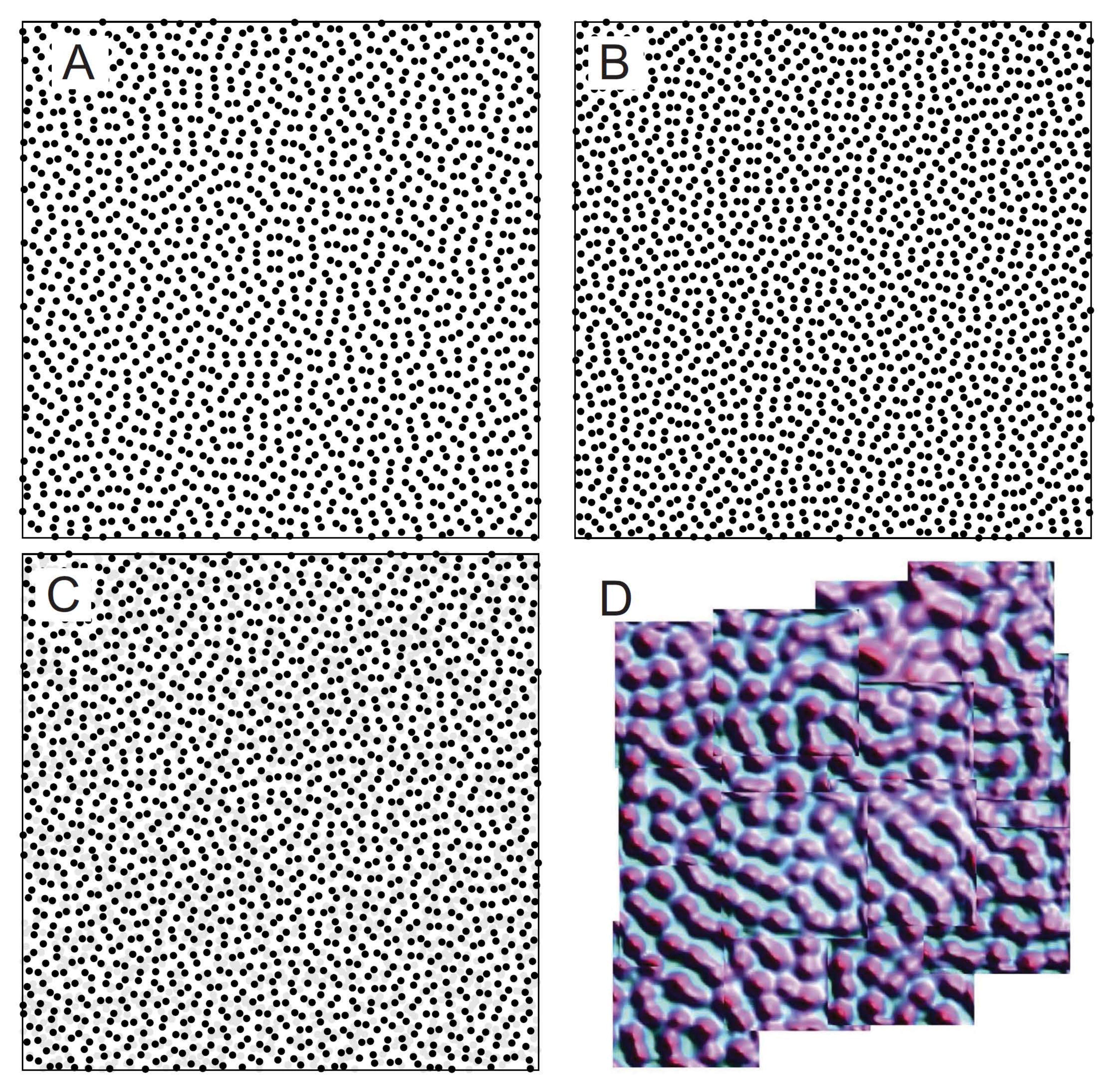}
\end{center}
\vspace{-0.5cm}
\caption{
Vortex patterns
in case of frustrated vortex-vortex interaction
for $\beta=0.6$
and
for varying
pinning strength and
number of vortices
per simulation cell:
$f_{p}=0$, $N_{v}=3600$ (A),
$f_{p}=0$, $N_{v}=4225$ (B), and
$f_{p}=0.3$, $N_{v}=3600$ (C).
Panel (D)
shows scanning Hall probe image ($\approx 50 \times 50 \ \mu$m$^{2}$) of the vortex distribution in a 160 nm thick superconducting MgB$_{2}$ film at T = 1.7 K and magnetic field of 2.8 Oe \cite{bending}.
}
\label{frb06np}
\end{figure}

In Fig.~\ref{frb06np} examples of vortex patterns are shown for $\beta=0.6$ and varying vortex density and the pinning strength.
As described above, frustration for high enough densities (for low densities, vortices form a hexagonal lattice which is not shown) leads to the formation of vortex dimers and four-vortex stripes (Fig.~\ref{frb06np}(A)), the morphology that appears to be robust with respect to increasing the vortex density (Fig.~\ref{frb06np}(B)). 
These vortex patterns are a disordered mixture (liquid) of predominantly vortex dimers and four-vortex chains, with a small fraction of single vortices and three-vortex chains. 
Disorder, as expected, 
induces irregular elastic deformations (i.e., local stretching or squeezing) of vortex chains or eventually plastic deformations (i.e., breaking the chains apart). 
Therefore, regular chains (four-vortex or longer with equidistant vortex distribution inside) either melt to disordered vortex dimers or turn to non-equidistant vortex chains consisting of vortex dimers rather than individual vortices [Fig.~\ref{frb06np}(C)]. 
The revealed features are 
similar to those seen in 
scanning Hall probe image of the vortex distribution in a 160 nm thick superconducting MgB$_{2}$ film measured at T = 1.7 K and magnetic field of 2.8 Oe \cite{bending}.
This analysis allows us to suggest that the experimental distributions can be understood in terms of the interplay of frustrated non-monotonic repulsive-attractive vortex-vortex interaction and pinning in MgB$_{2}$ films.

\section{Conclusions}

We analyzed effects related to 
vortex-core
deformations in the vicinity of other vortices.
These deformations 
result~\cite{coredeformation}
in the appearance of an additional attractive term in the overall repulsive vortex-vortex interaction that leads to the formation of various vortex patterns,  like vortex stripes, labyrinths, deformed lattices, etc., typical for systems with non-monotonic repulsive-attractive interaction.
However, real physical systems as, e.g., vortex matter in superconductors or in Bose-Einstein condensates, show more complex patterns which are due to the interplay of non-monotonic interaction (like in two-band or low-$\kappa$ superconductors) and other factors among which are disorder (e.g., due to random pinning which is inevitably present in superconductors) and eventually frustration in the vortex-vortex interaction which arises from the elongation of vortex cores in the direction of the closest neighbor. 

Using molecular-dynamics simulations, the effects related to the presence of random pinning and frustration in the vortex-vortex interaction have been investigated in detail.

First, we analyzed the zero-pinning case, 
including non-monotonic interactions but no orientational frustration.
We revealed the following phases (or ``morphologies'') and their sequence for increasing vortex density: 
(i) a hexagonal vortex lattice,
(ii) a liquid of vortex dimers,
(iii) short stripes,
(iv) long stripes,
(v) interconnected stripes and labyrinths, and
(vi) kagom\'{e} lattices.

Next, we introduced a weak random pinning and increased its strength.
Our analysis showed that all the above patterns, except for the hexagonal vortex lattice, appear to be rather sensitive to imperfections in the system.
In particular, we demonstrated that random pinning in the system leads to disordering and shortening of long vortex stripes (obtained in an ideal pinning-free system) and breaking the junctions between the stripes in the labyrinth-like configurations.
As a result, the obtained patterns are a mixture of short branching stripes and individual vortices or vortex dimers.
When the effect of vortex core deformations is strong 
(and, therefore, the attractive component in the vortex-vortex interaction is also strong) 
the formation of double stripes is observed, 
and these double stripes 
are also deformed and fragmented due to random pinning.
We compared the simulated vortex patterns with the experimental
patterns observed in MgB$_{2}$ films.
This comparison of morphologies allows us to identify the experimental images as the result of the interplay of the non-monotonic repulsive-attractive vortex-vortex interaction and pinning.

Further, we analyzed the effect of frustration in the vortex-vortex interaction.
We demonstrated that for high enough vortex densities the formation of vortex dimers and four-vortex stripes is favored.
The resulting vortex patterns are a disordered mixture (liquid) of predominantly vortex dimers and four-vortex chains, with a small fraction of single vortices and three-vortex chains.
Additional disorder facilitates the breaking apart of vortex chains, due to elastic deformations of the stripes. 
As a result, regular chains (four-vortex or longer with equidistant vortex distribution inside) either melt to disordered vortex dimers or turn to non-equidistant vortex chains consisting of vortex dimers rather than individual vortices.

Our findings can also be applicable to other vortex systems where the effects related to non-monotonic vortex-vortex interaction and frustration are applicable, like multi-band and low-$\kappa$ superconductors and Bose-Einstein condensates.

\smallskip

\section{Acknowledgments}

We acknowledge 
fruitful discussions with E.~Babaev and V.~Gladilin. 
This work is partially supported by 
the Natural Science Foundation of Jiangsu Province 
(Grant No.~BK20150595), 
the National Natural Science Foundation of China 
(Grants No.~NSFC-U1432135, No.~11611140101, and No.~11674054), 
the ``Odysseus'' program of the Flemish Government 
and Flemish 
Research 
Foundation (FWO-Vl), 
the Flemish Research Foundation (through Projects G.0115.12N, G.0119.12N, G.0122.12N, and G.0429.15N), 
the Research Fund of the University of Antwerp, 
RIKEN iTHES Project, 
the MURI Center for Dynamic Magneto-Optics 
via the AFOSR Award No.~FA9550-14-1-0040, 
the IMPACT program of JST, 
a Grant-in-Aid for Scientific Research (A), 
the Japan Society for the Promotion of Science (KAKENHI), 
CREST, and a grant from the John Templeton Foundation.


\begin{references}

\bibitem{aftalion} 
A.~Aftalion, 
{\it Vortices in Bose-Einstein Condensates}, 
(Springer, 2006). 

\bibitem{Wambaugh}
J.~F.~Wambaugh, C.~Reichhardt, C.~J.~Olson, F.~Marchesoni, and F.~Nori, 
{\it Superconducting Fluxon Pumps and Lenses}, 
Phys. Rev. Lett. {\bf 83}, 5106 (1999). 

\bibitem{vvmratchet} 
J.~Cuppens, G.~W.~Ataklti, V.~V.~Moshchalkov, A.~V.~Silhanek, 
J.~Van de Vondel, C.~C.~de Souza Silva, R.~M.~da Silva, and 
J.~A.~Aguiar
{\it Current-induced vortex trapping in asymmetric toothed channels},  Phys. Rev. B {\bf 84}, 184507 (2011). 

\bibitem{linratchet}
N.~S.~Lin, T.~W.~Heitmann, K.~Yu, B.~L.~T.~Plourde, and V.~R.~Misko, 
{\it Rectification of vortex motion in a circular ratchet channel}, 
Phys. Rev. B {\bf 84}, 144511 (2011). 

\bibitem{wepenprl}
V.~Misko, S.~Savel'ev, and F.~Nori,
{\it Critical currents in quasiperiodic pinning arrays: Chains and Penrose lattice},
Phys. Rev. Lett. {\bf 95}, 177007 (2005).

\bibitem{wepenprb}
V.~R.~Misko, S.~Savel'ev, and F.~Nori,
{\it Enhancement of the critical current in quasiperiodic pinning arrays: One-dimensional chains and Penrose lattices},
Phys. Rev. B {\bf 74}, 024522 (2006).

\bibitem{pendieterprl}
M.~Kemmler, C.~G\"{u}rlich, A.~Sterck, H.~P\"{o}hler, M.~Neuhaus, 
M.~Siegel, R.~Kleiner, and D.~Koelle, 
{\it Commensurability Effects in Superconducting Nb Films with Quasiperiodic Pinning Arrays}, 
Phys. Rev. Lett. {\bf 97}, 147003 (2006). 

\bibitem{pensilhanekapl}
A.~V.~Silhanek, W.~Gillijns, V.~V.~Moshchalkov, B.~Y.~Zhu, J.~Moonens,  and L.~H.~A.~Leunissen, 
{\it Enhanced pinning and proliferation of matching effects in a superconducting film with a Penrose array of magnetic dots}, 
Appl. Phys. Lett. {\bf 89}, 152507 (2006). 

\bibitem{penvvmprl}
R.~B.~G.~Kramer, A.~V.~Silhanek, J.~Van de Vondel, B.~Raes, and 
V.~V.~Moshchalkov, 
{\it Symmetry-Induced Giant Vortex State in a Superconducting Pb Film with a Fivefold Penrose Array of Magnetic Pinning Centers}, 
Phys. Rev. Lett. {\bf 103}, 067007 (2009). 

\bibitem{babaevPRB2014}
Q.~Meng, C.~N.~Varney, H.~Fangohr, and E.~Babaev,
{\it Honeycomb, square, and kagome vortex lattices in superconducting systems with multiscale intervortex interactions},
Phys. Rev. B {\bf 90}, 020509(R) (2014).


\bibitem{Babaev2005}
E. Babaev and M. Speight,
{\it Semi-Meissner state and neither type-I nor type-II superconductivity in multicomponent superconductors},
Phys. Rev. B {\bf 72}, 180502 (2005).

\bibitem{VVM2009}
V. Moshchalkov, M. Menghini, T. Nishio, Q.~H. Chen, A.~V. Silhanek, V.~H. Dao, L.~F. Chibotaru, N.~D. Zhigadlo and J. Karpinski,
{\it Type-1.5 Superconductivity},
Phys. Rev. Lett. {\bf 102}, 117001  (2009).

\bibitem{Nishio2010}
T. Nishio, V.~H. Dao, Q. Chen, L.~F. Chibotaru, K. Kadowaki and V.~V.
  Moshchalkov,
{\it Scanning SQUID microscopy of vortex clusters in multiband superconductors},
Phys. Rev. B {\bf 81}, 020506  (2010).

\bibitem{Babaev2010}
E. Babaev, J. Carlstr\"om and M. Speight,
{\it Type-1.5 Superconducting State from an Intrinsic Proximity Effect in Two-Band Superconductors},
Phys. Rev. Lett. {\bf 105}, 067003 (2010).

\bibitem{silhanek}
J. Gutierrez, B. Raes, A. V. Silhanek, L. J. Li, N. D. Zhigadlo, J. Karpinski, J. Tempere, and V. V. Moshchalkov,
{\it Scanning Hall probe microscopy of unconventional vortex patterns in the two-gap MgB$_{2}$ superconductor},
Phys. Rev. B {\bf 85}, 094511 (2012).


\bibitem{Berezin} 
V. A. Berezin and V. A. Tulin, 
{\it Attraction–Repulsion Transition between Abrikosov Vortices and Josephson Vortices in the Strongly Layered Superconductor Bi$_{2}$Sr$_{2}$CaCu$_{2}$O$_{8}$}, 
Phys. Sol. State {\bf 42}, 415 (2000). 

\bibitem{BuzdinPRL} 
A. Buzdin and I. Baladi\'{e}
{\it Attraction between Pancake Vortices in the Crossing Lattices of Layered Superconductors}, 
Phys. Rev. Lett. {\bf 88}, 147002 (2002). 
 

\bibitem{BuzdinPRB} 
A. V. Samokhvalov, A. S. Mel'nikov, and A. I. Buzdin, 
{\it Attraction between pancake vortices and vortex molecule formation in the crossing lattices in thin films of layered superconductors}, 
Phys. Rev. B {\bf 85}, 184509 (2012). 
 

\bibitem{Auer} 
J. Auer and H. Ullmaier, 
{\it Magnetic Behavior of Type-II Superconductors with Small Ginzburg-Landau Parameters}, 
Phys. Rev. B {\bf 7}, 136 (1973). 


\bibitem{eilenberger}
G.~Eilenberger and H.~B\"{u}ttner, 
{\it The structure of single vortices in type II superconductors}, 
Z. Physik {\bf 224}, 335 (1969). 


\bibitem{Jacobs} 
L. Jacobs and C. Rebbi, 
{\it Interaction energy of superconducting vortices}, 
Phys. Rev. B {\bf 19}, 4486 (1979). 


\bibitem{brandt}
E. H. Brandt, 
{\it Microscopic theory of clean type II superconductors in the entire field-temperature plane}, 
phys. stat. sol. (b) {\bf 77}, 105 (1976). 
 

\bibitem{klein}
U. Klein, J. Rammer, and W. Pesch, 
{\it A simple interpolation method for the free energy of a type II superconductor in a mixed state}, 
J. Low Temp. Phys. {\bf 66}, 55 (1987). 
 

\bibitem{sluchanko}
N. Sluchanko, S. Gavrilkin, K. Mitsen, A. Kuznetsov, I. Sannikov, V. Glushkov, S. Demishev, A. Azarevich, A. Bogach, A. Lyashenko, et al. 
{\it Superconductivity in ZrB$_{12}$ and LuB$_{12}$ with Various Boron Isotopes}, 
J. Sup. Novel Mag. {\bf 26}, 1663 (2013). 
 

\bibitem{ge}
J.-Y. Ge, J. Gutierrez, A. Lyashchenko, V. Filipov, J. Li, and V. V. Moshchalkov, 
{\it Direct visualization of vortex pattern transition in ZrB$_{12}$ with Ginzburg-Landau parameter close to the dual point}, 
Phys. Rev. B {\bf 90}, 184511 (2014). 


\bibitem{prozorov} 
R. Prozorov, 
{\it Equilibrium Topology of the Intermediate State in Type-I Superconductors of Different Shapes}, 
Phys. Rev. Lett. {\bf 98}, 257001 (2007). 
 


\bibitem{babaevPRB2015}
J.~Garaud and E.~Babaev,
{\it Vortex chains due to nonpairwise interactions and field-induced phase transitions between states with different broken symmetry in superconductors with competing order parameters},
Phys. Rev. B {\bf 91}, 014510 (2015).

\bibitem{Vedmedenko2007}
E.~Y. Vedmedenko,
{\it Competing interactions and patterns in nanoworld}
(Wiley-VCH Verlag GmbH \& Co. KGaA, 2007).

\bibitem{Ball1999}
P.~Ball,
{\it The Self-Made Tapestry: Pattern Formation in Nature}
(Oxford Univ. Press, Oxford, U.K., 1999).

\bibitem{FNsci97}
F.~Nori,
{\it Intermittently flowing rivers of quantized magnetic flux},
Science {\bf 271}, 1373 (1996).

\bibitem{FNprb97}
C.~J.~Olson, C.~Reichhardt, and F.~Nori,
{\it Superconducting vortex avalanches, voltage bursts,
and vortex plastic flow: Effect of the microscopic pinning
landscape on the macroscopic properties},
Phys. Rev. B {\bf 56}, 6175 (1997).

\bibitem{FNprl98} 
C.~J.~Olson, C.~Reichhardt, and F.~Nori,
{\it Fractal networks, braiding channels, and voltage noise in intermittently flowing rivers of quantized magnetic flux}, 
Phys. Rev. Lett {\bf 80}, 2197 (1998). 

\bibitem{FNprl99} 
A.~P.~Mehta, C.~Reichhardt, C.~J.~Olson, and F.~Nori, 
{\it Topological invariants in microscopic transport on rough landscapes: Morphology and Horton analysis of river-like networks of vortices}, 
Phys. Rev. Lett. {\it 82}, 3641 (1999). 

\bibitem{kagome}
M.~F.~Laguna, C.~A.~Balseiro, D.~Dom\'{i}nguez, and F.~Nori,
{\it Vortex structure and dynamics in kagom\'{e} and triangular pinning potentials},
Phys. Rev. B {\bf 64}, 104505 (2001).

\bibitem{Zhao2012NJP}
H.~J.~Zhao, V.~R.~Misko and F.~M.~Peeters,
{\it Analysis of pattern formation in systems with competing range interactions},
New Journal of Physics {\bf 14}, 063032 (2012).

\bibitem{Zhao2013PRE}
H.~J.~Zhao, V.~R.~Misko and F.~M.~Peeters,
{\it Dynamics of self-organized driven particles with competing range interaction},
Phys. Rev. E {\bf 88}, 022914 (2013).

\bibitem{Xu} 
X.~B.~Xu, H.~Fangohr, M.~Gu, W.~Chen, Z.~H.~Wang, F.~Zhou, D.~Q.~Shi, and S.~X.~Dou, 
{\it Simulation of the phase diagram of magnetic vortices in two-dimensional
superconductors: evidence for vortex chain formation}, 
J. Phys.: Condens. Matter {\bf 26}, 115702 (2014). 

\bibitem{bending}
P. J. Curran, W. M. Desoky, M. V. Milosevic, A. Chaves, J.-B. Lalo\"{e}, J. S. Moodera \& S. J. Bending,
{\it Spontaneous symmetry breaking in vortex systems with two repulsive lengthscales},
Scientific Reports {\bf 5}, 15569 (2015), DOI: 10.1038/srep15569. 

\bibitem{bookbec} 
{\it Emergent Nonlinear Phenomena in Bose-Einstein Condensates: Theory and Experiment}, 
Eds. P.~G.~Kevrekidis, D.~J.~Frantzeskakis, R.~Carretero-Gonz\'{a}lez 
(Springer, 2008). 

\bibitem{jtprl2012}
J.~Van~de~Vondel, V.~N.~Gladilin, A.~V.~Silhanek, W.~Gillijns, 
J.~Tempere, J.~T.~Devreese, and V.~V.~Moshchalkov, 
{\it Vortex Core Deformation and Stepper-Motor Ratchet Behavior in a Superconducting Aluminum Film Containing an Array of Holes}, 
Phys. Rev. Lett. {\bf 106}, 137003 (2011). 

\bibitem{jtpc2012} 
J.~Tempere, E.~Vermeyen, B.~Van~Duppen, 
{\it Skyrmion rows, vortex rows, and phase slip lines in sheared 
multi-component condensates}, 
Physica C {\bf 479}, 61 (2012). 

\bibitem{frankel}
T.~Frankel, 
{\it The Geometry of Physics: An Introduction} (2nd Ed.), 
(Cambridge, 2004). 

\bibitem{sadoc}
J.-F.~Sadoc and R.~Mosseri, 
Jean-Fran\c{c}ois Sadoc and R\'{e}mi Mosseri, 
{\it Geometrical Frustration},
(Cambridge, 1999). 

\bibitem{nelson}
D.~R.~Nelson, 
{\it Defects \& Geometry in Condensed Matter Physics} (2nd Ed.), (Cambridge, 2002). 

\bibitem{nelsonprb83}
D. R.~Nelson,
{\it Order, frustration, and defects in liquids and glasses},
Phys. Rev. B {\bf 28}, 5515 (1983).

\bibitem{steinhardt}
P. J.~Steinhardt, D. R.~Nelson and M.~Ronchetti,
{\it Icosahedral Bond Orientational Order in Supercolled Liquids},
Phys. Rev. Lett. {\bf 47}, 1297 (1981).

\bibitem{weht}
V.~Misko and F.~Nori,
{\it Magnetic flux pinning in superconductors with hyperbolic-tessellation arrays of pinning sites},
Phys. Rev. B {\bf 85}, 184506 (2012).

\bibitem{FNprl94}
R.~A.~Richardson, O.~Pla, and F.~Nori,
{\it Confirmation of the modified bean model from simulations
of superconducting vortices},
Phys. Rev. Lett. {\bf 72}, 1268 (1994).

\bibitem{walter2009}
W.~V.~Pogosov, V.~R.~Misko, H.~J.~Zhao, and F.~M.~Peeters,
{\it Collective vortex phases in periodic plus random pinning potential},
Phys. Rev. B {\bf 79}, 014504 (2009).

\bibitem{coredeformation} 
The interaction of a vortex with a repulsive spatially-extended object (like deformed vortex core) can be modeled by a set of repulsive forces. 
While at long distances these forces are indistinguishable and 
can be approximated by one arising from the front of the 
deformed core, this force weakens at short distances, and the 
contribution from the rear of the extended core comes into play. 
This situation is similar to the case of multi-scale vortex-vortex interaction~\cite{babaevPRB2014} when a transition between overall repulsive force profiles results in the appearance of a threshold which can be described as a sum of a repulsive force and additional weak attractive force. 

\bibitem{overlapping} 
This is clear from the following simple consideration. 
In good type-II superconductors, where $\lambda \gg \xi$, vortex cores are well separated for fields below $H_{c2}$ and magnetic cores may overlap which provides intervortex repulsion and stability of the Abrikosov lattice. 
In good type-I superconductors, where $\lambda \ll \xi$, vortex cores may overlap while magnetic cores remain well separated which result in unstable vortex patterns and their collapse to the mixed state of type-I superconductors. 


\bibitem{gladilin}
V. N. Gladilin, J. Ge, J. Gutierrez, M. Timmermans, J. Van de Vondel, J . Tempere, J. T. Devreese, and V. V. Moshchalkov, 
{\it Vortices in a wedge made of a type-I superconductor}, 
New J. Phys. {\bf 17}, 063032 (2015). 


\bibitem{walter2010}
W.~V.~Pogosov, H.~J.~Zhao, V.~R.~Misko, and F.~M.~Peeters,
{\it Kink-antikink vortex transfer in periodic-plus-random pinning potential: Theoretical analysis and numerical experiments},
Phys. Rev. B {\bf 81}, 024513 (2010).


\end{references}
\end{document}